\documentclass[a4paper,fleqn,usenatbib]{mnras}

\usepackage{newtxtext,newtxmath}
\usepackage[T1]{fontenc}
\usepackage{ae,aecompl}

\usepackage{graphicx}	% Including figure files
\usepackage{amsmath}	% Advanced maths commands
\usepackage{amssymb}	% Extra maths symbols

\title[Uncovering MPs in M30 using Washington Photometry.]{Uncovering Multiple Populations in NGC 7099 (M 30) using Washington Photometry}

\author[Frelijj et al.]{
H. Frelijj,$^{1}$\thanks{E-mail: hfrelijj@astro-udec.cl}
D. Geisler,$^{1}$
J. Cummings,$^{2}$
R. E. Cohen,$^{1,6}$
\newauthor
F. Mauro,$^{4,1,5}$
C. Munoz,$^{1,3}$
S. Villanova,$^{1}$
and B. Tang$^{3}$
\\
$^{1}$Departamento de Astronom\'ia, Casilla 160-C, Universidad de Concepci\'on, Chile\\
$^{2}$Center for Astrophysical Sciences, Johns Hopkins University, 3400 North Charles Street, Baltimore, MD 21218, USA\\
$^{3}$European Southern Observatory, Casilla 19001, Santiago, Chile\\
$^{4}$Millennium Institute of Astrophysics, Chile\\
$^{5}$Universidad Cat\'olica del Norte, Angamos 0610, Antofagasta, Chile\\
$^{6}$Space Telescope Science Institute, 3700 San Martin Drive, Baltimore, MD 21218, USA 
}

\date{Accepted XXX. Received YYY; in original form ZZZ}

\pubyear{2017}

\begin{document}
\label{firstpage}
\pagerange{\pageref{firstpage}--\pageref{lastpage}}
\maketitle

\begin{abstract}
Over the last decade, the classical definition of Globular Clusters (GCs) as simple stellar populations was revolutionized due to the discovery of "Multiple Populations" (MPs). However, our knowledge of this phenomenon and its characteristics is still lacking greatly observationally, and there is currently no scenario that adequately explains its origin. It is therefore important to study as many GCs as possible to characterize whether or not they have MPs, and determine their detailed behavior to enlighten formation scenarios, using a wide range of techniques. The Washington photometric system has proved to be useful to find MPs thanks mainly to the UV-sensitivity and high efficiency of the C filter. We search for MPs in the Galactic GC NGC 7099 (M30), the second GC being searched for MPs using this system. We obtained photometric data using the Swope 1m telescope at Las Campanas Observatory, as well as the 4m SOAR facility. Our reduction procedure included addstar experiments to properly assess photometric errors. We find a clear signal of MPs based on an intrinsically wide color spread on the RGB, in particular due to a relatively small fraction of stars significantly bluer than the main RGB locus. These stars should correspond to so-called first generation stars, which we estimate to be roughly 15\% of the total. However, we find these first-generation stars to be more spatially concentrated than their second generation counterparts, which is the opposite to the general trend found in other clusters. We briefly discuss possible explanations for this phenomenon. 
\end{abstract}


%%%%%%%%%%%%%%%%%%%%%%%%%%%%%%%%%%%%%%%%%%%%%%%%%%

%%%%%%%%%%%%%%%%% BODY OF PAPER %%%%%%%%%%%%%%%%%%

\section{Introduction}

 GCs have traditionally been considered as Simple Stellar Populations (SSPs), implying they originated in a single star formation event, with all stars having the same age and initial chemical content. However, a variety of observations, both spectroscopic and photometric,
over the past decade have revealed that the majority, if not all, GCs instead host "Multiple Populations". These populations contain stars with an intrinsic variation in certain properties, suggesting a more complex formation than previously believed. Due to the revolutionary work of \citet{Carretta2009}, involving observations of 1958 stars in 19 Galactic GCs using the high-resolution multiobject FLAMES spectrograph on the VLT, the most widely recognized symptom of MPs is an anticorrelation in the abundances of the light elements Na and O. All of the GCs surveyed in their study displayed this anticorrelation. Subsequent studies showed this behavior in a number of other GCs as well. However, an exception is found in at least one genuine old, massive GC that apparently does not possess MPs: Ruprecht 106 \citep{Villa2013}. 
 
 High resolution spectroscopic studies are a very powerful diagnostic for studying the MP phenomenon because they yield detailed abundances for a number of different species with a wide range in their nucleosynthetic origin. However, they are limited by the number of bright stars for which high resolution spectra of sufficient S/N can be obtained.
 
 Photometry is another, complementary way to search for MPs. While it cannot provide the detailed abundances of spectroscopy, it allows the measurement of a much larger sample of stars simultaneously and to much fainter absolute magnitudes. It has been known for more than 40 years that the most massive Galactic GC, Omega Cen, possesses a giant branch that is much wider than the photometric errors \citep{Cannon1973}. Subsequent modern HST studies of this intriguing cluster \citep{Bedin2004} demonstrated that the CMD showed multiple sequences in not only the RGB but the SGB and MS as well. Most recently, \citet{Piotto2015} have used a combination of three filters with the UVIS/WFC3 instrument on board the Hubble Space Telescope (HST) to examine MPs in $\sim60$ Galactic GCs. They find strong evidence of MPs in all of their sample through a significant spread or even a split in not only the Red Giant Branch (RGB) but also in the Sub Giant Branch (SGB) or even the Main Sequence (MS).

 A number of studies now point to the fact that certain filters are better than others at identifying MPs. In particular, filters covering the blue to near-UV portion of the spectrum are most effective, as revealed by e.g. \citet{Marino2008} and \citet{Han2009}. \citet{Sbor2011} investigated this in detail. They produced synthetic spectra of two otherwise identical GC giants, one is a so-called First Generation star with normal chemical abundances of the light elements and the other is a Second Generation star with enhanced He, N and Na and depleted C and O, as observed in many spectroscopic studies, and investigated the effects on the flux in various filters. A majority of the differences between the two spectra are related to the various CN, CH, NH and OH bands. These are concentrated in the blue-UV part of the spectrum, from the CH G-band at 4300 \AA \ to the OH band at 2800 \AA. Although there are additional features such as CN bands in the red and near-IR, this is also where the flux of cool giants peaks, minimizing the effects of any differences, while the above-mentioned bands produce proportionately larger differences in the much-reduced blue-UV flux of cool giants. Thus filters in this region are extremely effective MP tracers. Particularly effective at uncovering MPs are the so-called "magic trio" of HST UVIS/WFC3 filters: F275W, F336W and F438W \citep{Piotto2015}. Flux differences amounting to several tenths of a magnitude are typically found between First Generation(FG) and Second Generation(SG) RGB stars. The combination of MP sensitivity and HST imaging allows one to make very detailed "chromosome maps" tracking the definitive details of MPs within a GC. This group is conducting a UV legacy survey of Galactic GCs using this very powerful technique. They have uncovered a bewildering variety of MP behavior. The 57 GCs in their sample display 57 different color distributions! All of them possess MPs (although note that they do not include Ruprecht 106 in their sample). Any GC formation scenario successfully explaining the full range of observations will need to be complex indeed. However, in general \citet{Milone2017} find that most GCs separate into 2 main groups in their chromosome mapping, which are confirmed by spectra to correspond to FG and SG stars. However, some GCs display an intrinsic range of colors even within one of these groups, so that even FG stars are not necessarily born as SSPs! They also find that the percentage of FG stars within a GC varies widely, and shows a significant correlation only with GC luminosity and mass and not with any other global properties.

Even more powerful diagnostics can be achieved when one combines both excellent photometry and spectroscopy. A case in point is that of NGC 2808, where \citet{Milone2015} combine their magic trio data with existing FLAMES data for a relatively large number of giants. They find at least 5 separate populations in the chromosome diagram, which map cleanly into different groups in the Na-O anticorrelation. The question of course arises as to whether the various populations are indeed genetically discrete or instead the result of continuous variation. The answer is not yet clear but the best evidence suggests both possibilities may occur.
 
 As discussed above, the blue-UV part of the spectrum is especially sensitive to MPs. We recently realized that a particularly useful filter should be the Washington system's C filter \citep{Canterna1976}. This broadband system was designed originally to derive a photometric temperature (from the $T_1$ and $T_2$ filters, very similar to $(RI)_{KC}$.), as well as a metallicity index (from the M filter) for G and K giants. However, at the time, CN and CH variations were being discovered in GCs and it was felt prudent to include another filter that would be sensitive to such variations independent from metallicity effects, and thus the C filter was added. This filter goes from the atmospheric cutoff at around 3300 \AA \ to beyond the G-band, thus covering all of the spectral range most affected by MPs visible from the ground. The designations of the Washington filters, $CMT_1T_2$, are appropriate, given their Carbon (CN/CH abundance), Metallicity, and Temperature sensitivities. Indeed, the Washington C filter was the first explicitly designed to be sensitive to MPs, although they were not recognized as such at the time. The C filter is much broader (>1000 \AA) than other alternatives like Johnson U, or Sloan or Stromgren u. In addition, it is centered somewhat redder than these filters and thus is less affected by both reddening as well as atmospheric extinction. We thus investigated the sensitivity of the C filter in discriminating the presence of MPs. The test case was NGC 1851, a typical GC with well studied MPs. In order to demonstrate its efficiency, we opted to obtain the data with only a 1m class telescope from the ground. We observed the cluster for a total of 3h in the C filter on the LCO 1m Swope telescope, also observing in the $R_{KC} I_{KC}$($T_1T_2$) filters. Despite the small aperture relatively large (0.44'') pixels and mediocre seeing (1-2''), the Washington C vs. $C − T_1$ CMD \citep{Cummings2014} clearly shows a very broad RGB, much wider than the photometric errors. In addition, the reddest stars along the RGB are indeed those that are known to be N and Na rich from spectra, while the bluest are N and/or Na poor \citep{Cummings2017}. Thus, the C filter is indeed very effective in discriminating MPs, despite its broadband nature, and very efficiently, as expected. This is very encouraging news. Indeed, the WFC3 on board HST includes a C filter, F390W, which should be very effective in studying MPs in very distant systems, e.g. M31 GCs, which cannot be reached by the much narrower band magic trio filters. Furthermore, HST will not last much longer. When it does deorbit, we will be left uv-blind, with no capability of observing below the atmospheric cutoff and thus no possibility of observing over 1/2 of the magic trio spectral range. However, the C filter was designed for ground-based work. Thus, in the very near future, this will become perhaps the most effective photometric means of studying MPs.  
 
Despite its long history, our understanding of MPs, both observationally and theoretically, is in its infancy. We badly need more input from both large scale photometric and spectroscopic programs of star clusters covering a wide variety of ages, masses, metallicities and galactic environments in order to help constrain the new ideas regarding their nature and origin that are desperately needed to properly understand them.
   
 Our goal in this paper is to search for MPs in another GC, NGC 7099 (M30), using the Washington system in order to further investigate the utility of the system to uncover MPs.
 NGC 7099 is a very metal poor cluster([Fe/H]= -2.27 dex) \citep[updated as in 2010]{Harris2010} situated in the galactic halo at 8.1 Kpc \citep[updated as in 2010]{Harris2010} from our Sun, with coordinates(epoch J2000): RA: $21^h 40^m 22.12^s$, Dec: -23$^\circ$ 10' 47.5''. Its absolute V magnitude is -7.45, near the peak of the GC luminosity function, and it has a foreground reddening of only E(B-V)=0.03 \citep[updated as in 2010]{Harris2010}.
 It has an estimated age of 12.9 Gyrs \citep{Forbes2010} and a mass of $1.6\times10^6 M_\odot$ \citep{Vande2009}.

 Previous work on this cluster searching for MPs found spectroscopically a [Na-O] anticorrelation showing two distinct populations \citep{Carretta2009}. Photometrically, \citet{Piotto2015} and \citet{Milone2017} have also found MPs in the wide RGB displayed in their magic filter data for this GC.

 This paper is organized as follows: In section 2, we discuss our observations and reduction. In section 3, we discuss how we "optimized" the confidence of our data, first by conducting addstar experiments to determine whether or not our internal errors are good estimates of the real errors and secondly culling the data by eliminating stars with large errors and/or lying in the crowded central regions of the cluster. In section 4 we search for and indeed find MPs via CMD analysis. We then investigate their relative numbers and radial distributions. In section 5 we summarize our results and discuss possible scenarios to explain the unexpected result we find that the putative FG stars are more centrally concentrated than their SG counterparts. 

\section{The Data}
\subsection{The Observations}
 The data consist of 26 images of which 6 were obtained with the 4m SOAR telescope on Cerro Pachon, Chile in 2014 and 20 were obtained with the 1-meter Swope telescope from Las Campanas Observatory, Chile. Fourteen of these latter images are from 2013 and 6 are from 2011. The SOAR detector (SOI) consists of a total of 4096 x 4096 pixels at 0.1534''/pix (0.0767 ''/pix binned 2x2) and a field of view of 5.26 x 5.26 arc minutes, divided into two CCDs with two amplifiers each resulting in 4 columns of 1024x4096 pixels. 
 The Swope telescope worked with only one CCD (SiTe3) of 2048x3150 pixels at 0.435 ''/pix and a field of view of 14.9 x 22.8 arc minutes.
 The filters used for this work were the Washington C filter \citep{Canterna1976}, and the filters $R_{KC}$ and $I_{KC}$ in replacement of the Washington filters $T_1$ and $T_2$. \citet{Geisler1996} demonstrated that the $R_{KC}$ filter is photometrically appropriate but more efficient substitute for $T_1$. Additionally, the $T_2$ filter is almost identical to $I_{KC}$ \citep{Canterna1976,Geisler1996}.
 We obtained 5 images each in R and I and 10 in C from the Swope, while only C data were obtained with the SOAR telescope, for a total of 16 in C, which is the crucial filter for detecting MPs but also the most difficult in which to obtain high S/N, especially for red giants.
\newline

 Table \ref{tab:exposures} Details of the exposures:
{%
\newcommand{\mc}[3]{\multicolumn{#1}{#2}{#3}}
\begin{center}
\label{tab:exposures}
\begin{tabular}{|l|l|l|}\hline
 & \mc{1}{c|}{Swope} & \mc{1}{c|}{SOAR}\\\hline
C & 1(30s), 3(300s), 6(1200s) & 4(10s), 2(300s)\\
R & 1(10s), 1(100s), 3(400s) & \mc{1}{c|}{-}\\
I & 1(10s), 1(300s), 3(1200s) & \mc{1}{c|}{-}\\\hline
\end{tabular}

\end{center}
}%

\vspace{4mm}
The 2011 Swope images have a FWHM of 1.1''-2.2'' and an airmass of 1.19-1.41, the 2013 Swope images have a FWHM of 0.9''-1.7'' and an airmass of 1.01-1.12, while the 2014 SOAR images have a FWHM of 1.1'' and an airmass of 1.01.
 The first night of the 2013 observations was deemed photometric via visual estimate of the sky conditions and we observed a number of standard fields from \citet{Geisler1996} which were later used to calibrate all of the data. Subsequent detailed reduction and analysis showed that this night was indeed photometric (see below).

\subsection{Processing and Reduction}
 For the processing part the program used was IRAF \footnote{IRAF is distributed by the National Optical Astronomy Observatory, which is operated by the Association of Universities for Research in Astronomy (AURA) under a cooperative agreement with the National Science Foundation.} and its standard tasks like ccdproc. For the SOAR images, since there were gaps between the chips, we decided to work with each amplifier as a separate image. 
 
  For all the Swope images we applied a nonlinearity correction from \citet{Hamuy2006}. The photometry was performed by the DAOPHOT suite of programs \citep{Stetson1987}, as incorporated into the IRAF environment. 
  
  The cluster is crowded, even in the SOAR images, demanding profile-fitting photometry for the best results. We derived the PSF using 150-200 isolated, bright stars in each image, first with an initial calculation by the DAOPHOT task psf, followed by a manual refinement by eye, then a more accurate refinement with a Fortran program that generates the PSF for all PSF stars and subtracts out their neighbors between iterations, and a final refinement using again the DAOPHOT task psf. After that, with a good PSF in hand, we performed the usual 3 passes through find, phot and allstar. Finally, once all images were ALLSTARed, ALLFRAME was applied to all the images simultaneously, allowing the most precise photometry \citep{Stetson1994}. Afterwards, the aperture correction was made by comparing the psf photometry of psf stars to their aperture photometry, correcting out to a radius of 17.5''.

  DAOMATCH and DAOMASTER were used to match all images from one filter to get a robust intensity-weighted mean instrumental magnitude, using a medium exposure as a reference image since it maximizes the number of stars in common with both short and long exposures, facilitating the match. 
  Both programs were used again to generate a full catalog with all the stars found in at least 2 of the 3 filters. The R filter was used as reference filter since its wavelength response lies between the C and I filters and because it produces the deepest images. 
  
 The instrumental magnitudes were transformed to the standard Washington system using the standard star observations obtained in 2013. The night did indeed prove to be photometric. We got an RMS of 0.038(C), 0.022(R) and 0.027(I) that we obtained from fitting the observations to the standard system using 63 standard stars for C, 65 for R and 74 for I, which covered a large range in color and airmass.  

  The final standardized photometry is given in Table \ref{tab:grandmaster}. After each magnitude there are two errors (i.e. eX and dX where X is a filter). The first is a statistical assessment of the psf-fitting error for each detection of a star returned by Allframe, in which the reported error is the weighted mean of the errors of the individual detections, where the weight is inversely proportional to the individual error. The second error measures the variation in magnitude (dispersion) of the various independent detections. Note that observations in the R and I filters have been transformed to $T_1$ and $T_2$ magnitudes. 
\begin{table*}
\centering
\begin{tabular}{lllllllllllllll}\hline
ID & X & Y & C & eC & dC & T1 & eT1 & dT1 & T2 & eT2 & dT2 & nC & nT1 & nT2\\\hline
5023 & 1121.36 & 1423.45 & 21.050 & 0.013 & 0.028 & 20.123 & 0.026 & 0.043 & 19.684 & 0.024 & 0.053 & 9 & 3 & 3\\
5034 & 1223.40 & 1423.21 & 19.388 & 0.005 & 0.025 & 18.736 & 0.017 & 0.007 & 18.339 & 0.022 & 0.067 & 15 & 3 & 3\\   
5059 & 775.29 & 1424.86 & 20.300 & 0.006 & 0.048 & 19.549 & 0.022 & 0.087 & 18.995 & 0.027 & 0.052 & 15 & 3 & 3      
\end{tabular}
\caption{\label{tab:grandmaster}The columns are: ID, X and Y coordinates (in px), magnitude, psf-fitting error and dispersion in C, $T_1$ and $T_2$, and the number of frames where the star was detected in C, $T_1$ and $T_2$.}
\end{table*} 

\section{Analysis}
\subsection{Addstar experiments}
In order to determine the best estimate of the true internal photometric errors, we performed Addstar experiments using the eponymous DAOPHOT task. First we added random fake stars (in magnitude and position) to one representative Swope frame for each filter. We then performed exactly the same photometry process as described above and then determined the difference in magnitude between the input and output values. We take this difference as our best estimate of the real internal photometric error (neglecting errors in the transformation, which will affect all stars equally). For this test, we added no more than $10\% $ of the original number of stars photometered in a frame and repeated the process 10 times, thereby measuring in the end the same number of added stars as actually in the original image.

\begin{figure}
\includegraphics[width=0.5\textwidth]{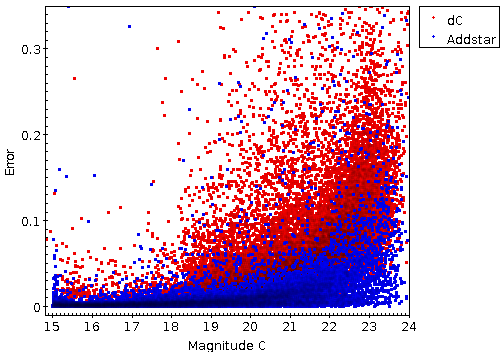}
\includegraphics[width=0.5\textwidth]{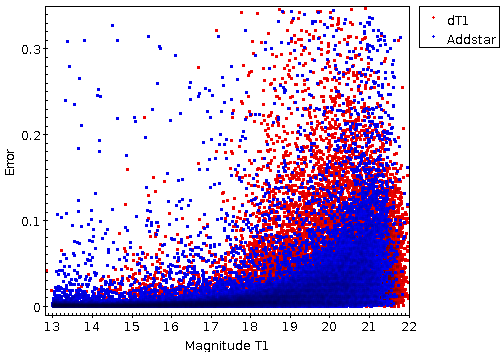}
\includegraphics[width=0.5\textwidth]{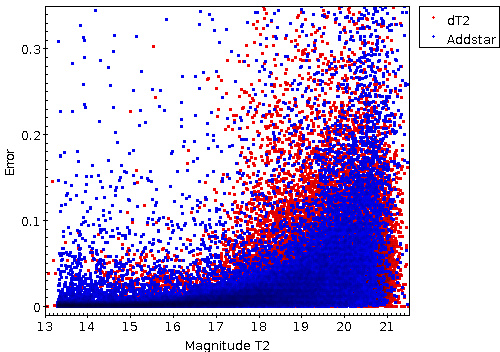}
\caption{\label{fig:addexp}Plots of magnitude vs two error assessments. The blue points are from the addstar experiments and the red points are the external errors measured from the dispersion in magnitude of each independent detection (dX in the table). The two are in reasonable agreement in general so we will use the external error as our photometric error.}
\end{figure}

In figure \ref{fig:addexp}, we compare the dispersion in the allframe magnitude (dX) error with the addstar error. 
The two error estimates are similar at all magnitudes in general. Clearly, for fainter magnitudes, the spread of both errors becomes greater. In general, the mean trends of both error assessments are in good agreement throughout the magnitude range, so we will therefore just use the allframe dispersion as our error estimate for each star.

\begin{figure}
\centering
\includegraphics[width=0.5\textwidth]{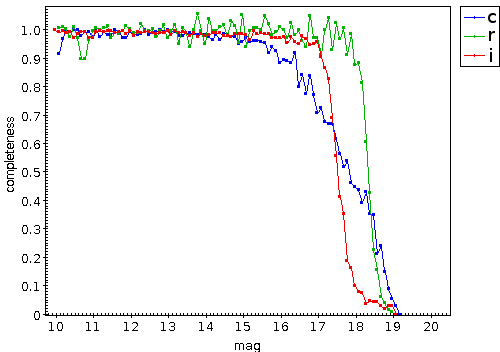}
\caption{\label{fig:magcomp}Plot of completeness as a function of magnitude based on the Addstar experiments.}
\end{figure}

We also plot in figure \ref{fig:magcomp} the completeness as a function of magnitude based on the Addstar experiments. The $50\% $ completeness level is found to be 17.9(c), 18.3(r) and 17.5(i), which translate roughly into 22(C), 22.4($T_1$) and 21.6($T_2$).

\begin{figure}
\centering
\includegraphics[width=0.5\textwidth]{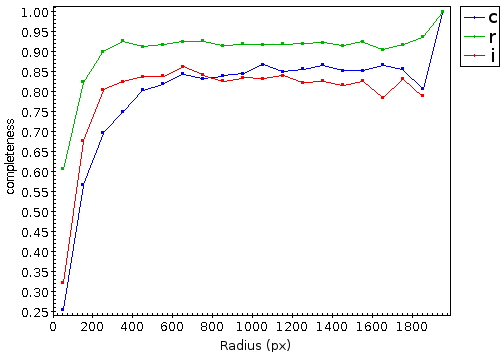}
\caption{\label{fig:radcomp}Plot of completeness as a function of radius based on the Addstar experiments.}
\end{figure}

Figure \ref{fig:radcomp} shows completeness as a function of radius. This plot shows that there were many fewer stars found inside a radius of 100 (Swope) pixels from the center of the cluster due to extreme crowding, suggesting caution should be used in including this area. However, an independent analysis was made later including the SOAR images, described below.  

 So we now have in hand definitive estimates of the errors and completeness and their variation with magnitude and radius which are required to make the best analysis concerning MPs.

\subsection{Final sample culling}

As expected, fainter stars generally have larger internal uncertainties and also a larger dispersion among different detections. However, there are also a relatively small number of brighter stars with small internal uncertainties but large dispersions, most likely due to crowding or undetected saturations. This fact, along with the comparison with the addstar errors, encourages us to prefer the external errors(dX) instead of the internal ones(eX), so from now on we refer to this value(dX) as the error. 

In order to optimize our sample for searching for MPs, we first removed all stars that were found in only 1 frame in a filter. We then removed all stars with an error larger than 0.1 and then placed the remaining stars in separate catalogues according to color. This eliminated very few bright stars and should have negligible effect on our main results, which are limited to brighter stars. 

We further cull our dataset to maximize the scientific return by first statistically correcting for field stars and investigating crowding. From our addstar experiment, we know that very few stars within 100px of the center on the Swope images were found and could be well measured. 
However, we can further investigate the radial behavior by adding the SOAR images.  We selected the range from 0-100px, the region with the greatest crowding problems seen in the Swope data. We next investigated crowding including the SOAR images as well and divided this inner region into annuli with a width of 10 (Swope) px. We found that beyond a radius of 50px the dispersion dropped significantly. We thus set 50 Swope px(21.7'') as the definitive inner useful radius. Similarly, by investigating the structure and dispersion of the various parts of the CMD we determined that field stars were most effectively removed by establishing a maximum cluster radius of 1000 Swope px(7.25'). The CMD beyond this radius shows no distinguishable RGB, which is the main focus of our investigation, so we lose a negligible fraction of these stars by using this limiting cluster radius. This value lies between NGC7099's half light radius: 1.03' \citep[updated as in 2010]{Harris2010}, and its tidal radius: 19' \citep[updated as in 2010]{Harris2010}.

\section{Revealing MPs}
\begin{figure*}
\includegraphics[width=0.495\textwidth]{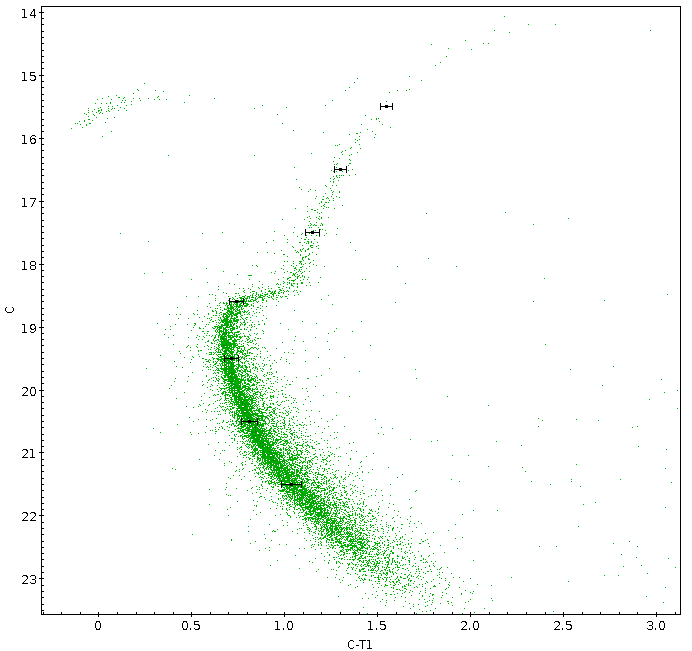}
\includegraphics[width=0.495\textwidth]{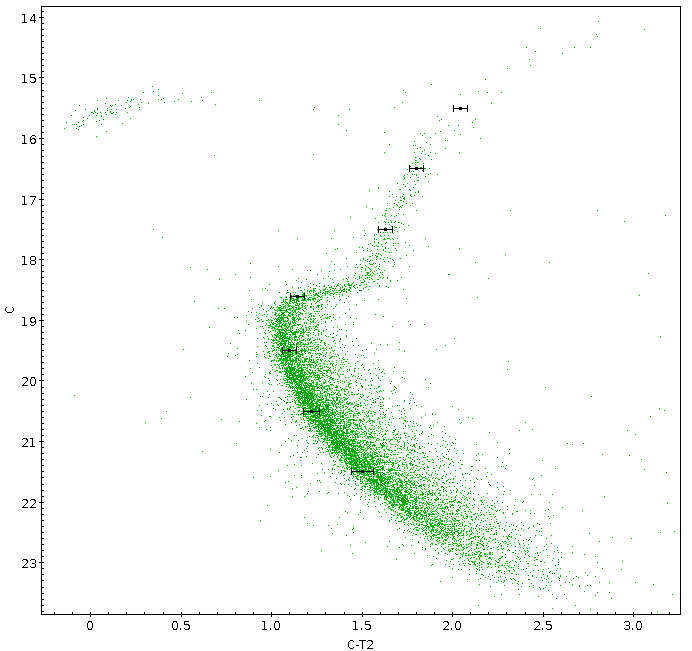}
\includegraphics[width=0.495\textwidth]{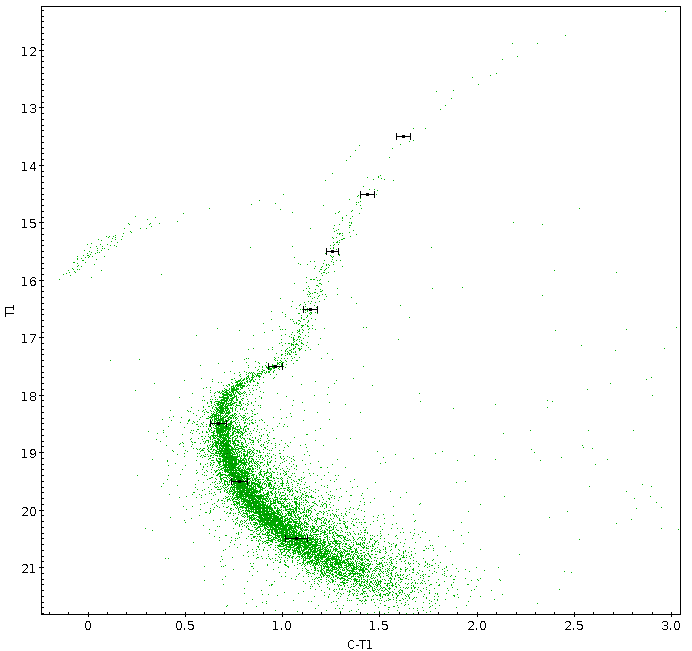}
\includegraphics[width=0.495\textwidth]{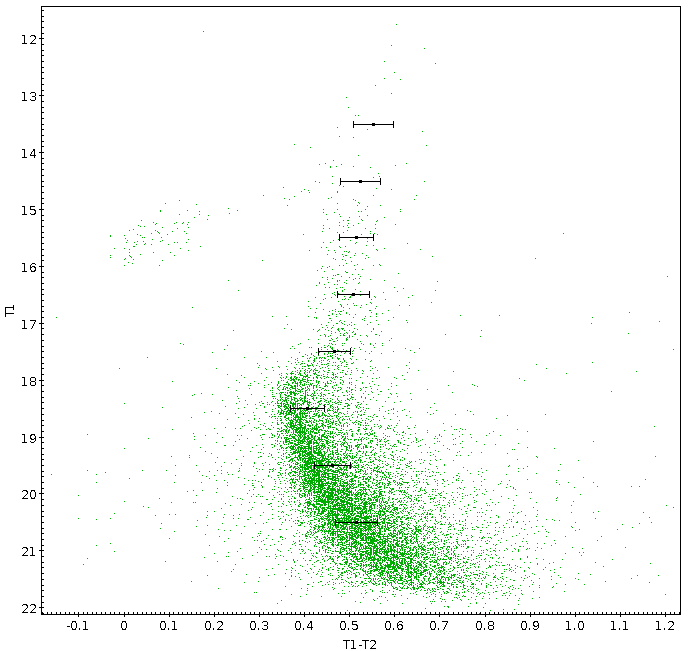}
\caption{\label{fig:CMD}The Washington CMDs. Errorbars corresponding to the mean error determined in 1 mag bins have been put along the primary sequences. All 3 CMDs including the CN/CH sensitive C filter show significant intrinsic color spreads along the mid to lower RGB, most notably by a scattering of stars to the blue of the main RGB locus. In contrast, the $T_1$ vs $ T_1-T_2$ CMD has a spread consistent with that expected from photometric errors alone. Note the different color scale for the latter CMD.}
\end{figure*}

With catalogues for each color containing only stars well measured and with small errors (a total of  $\sim$15000 in each catalogue), we now proceeded to make various CMDs by combining our 3 filters, including $T_1$ vs $C-T_1$,C vs $C-T_1$, C vs $C-T_2$, and $T_1$ vs $T_1-T_2$. Figure \ref{fig:CMD} shows these four CMDs. Black errorbars are shown representing the mean in each magnitude bin.
 {\bf{All 3 CMDs involving the C filter show a large color spread on the mid to lower RGB, larger than that expected from photometric errors alone, while the $T_1$ vs $T_1-T_2$ CMD color spread is consistent with no intrinsic variation.}}
 
 This is exactly the behaviour expected if NGC 7099 contains MPs, because the C filter is very sensitive to the differences in flux of the molecular bands of first and second generation stars, but $T_1$ and $T_2$ lack this sensitivity, making MPs indistinguishable.
 
 Since the reddening is very small (E(B-V)=0.03), differential reddening is assumed to be negligible. Actually $E(C-T_1)\sim 2 \times E(B-V)$, giving a value of $E(C-T_1)=0.06$, but this value is still very small and we will neglect any potential differential reddening. The negligible variable reddening means that our main results are not affected by this factor. 
 
 Note that, unlike the case in NGC 1851 and typical Washington photometry, our C photometric errors are actually smaller than those in the redder filters, especially $T_2$, due to the fact that we had a large number of long exposures with the Swope telescope as well as additional SOAR images, using a 4m telescope, which were only obtained for C.
 The net effect is to produce C errors which are quite small and clearly much smaller than the color spread we see in this part of the CMD. In particular, there are a number of stars that scatter well to the blue of the main RGB ridgeline, with essentially no counterparts to the red. All 3 CMDs involving the C filter have error bars demonstrating that this spread to the blue in the RGB is very unlikely due to errors. Note that the observational errorbars cover the main locus of the RGB in all cases, but in the diagrams with C they are too small to also include the stars to the blue of the RGB. The behavior in the different filters leads us to conclude that the cause of this spread is most likely due to MPs.

\begin{figure*}
\includegraphics[width=0.49\textwidth]{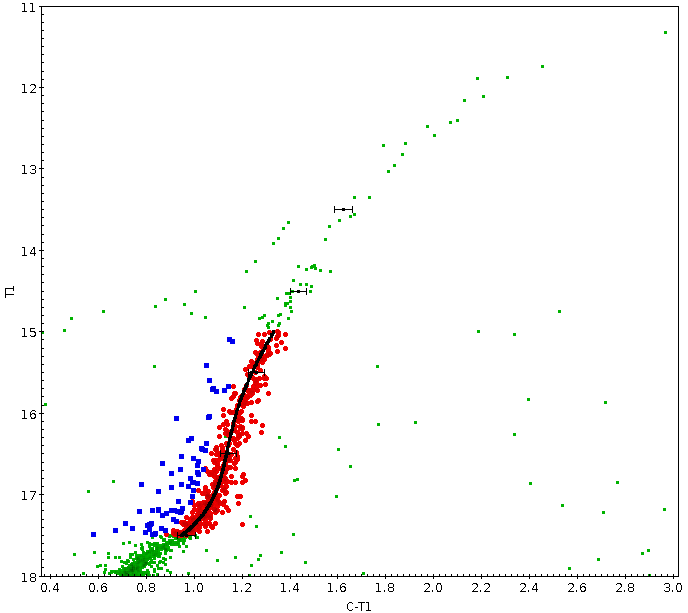}
\includegraphics[width=0.49\textwidth]{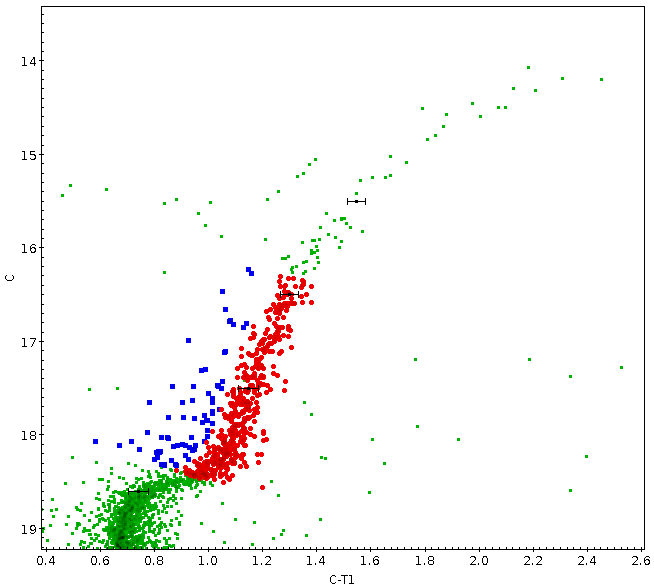}
\includegraphics[width=0.49\textwidth]{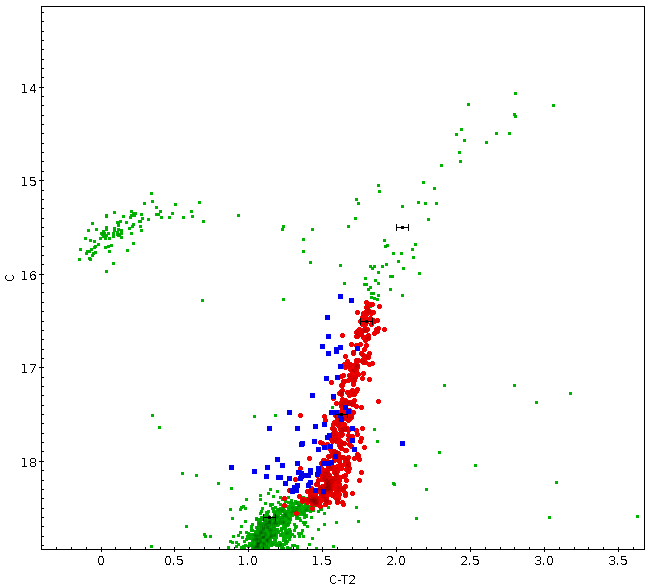}
\includegraphics[width=0.49\textwidth]{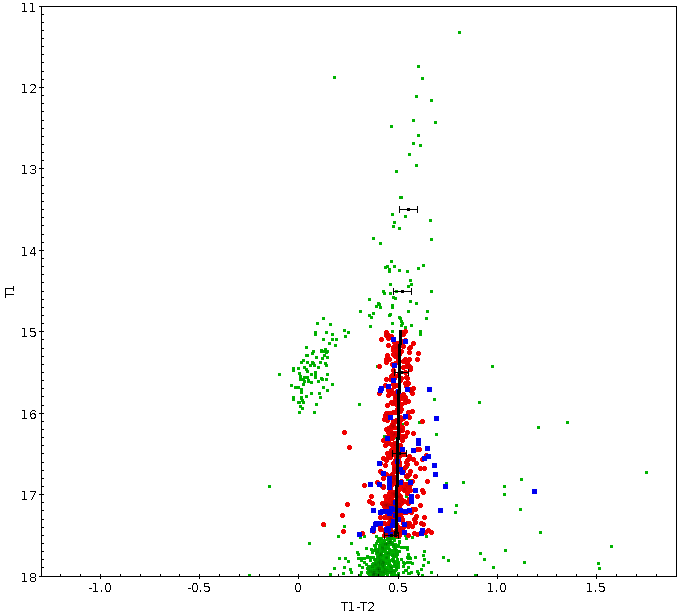}
\caption{\label{fig:RGB}Zoom in on the RGB. The two diagrams on the top and the bottom-left include the C filter while the bottom-right does not. We used the T1 vs C-T1 CMD to define blue and red stars according to their position along the lower RGB. These same stars appear as blue and red squares in the other 3 CMDs. In the two other CMDs involving the C filter, especially C vs. C-T1, the blue and red squares are generally also well separated. However, in the $T_1$ vs $T_1-T_2$ CMD, both sets of stars fall together along the main RGB. Fiducials are shown in $T_1$ vs $C-T_1$ and $T_1$ vs $T_1-T_2$}
\end{figure*}

Figure \ref{fig:RGB} shows a zoom in on the RGB. We color code in blue stars that lie to the blue of the main RGB locus in the T1 vs C-T1 CMD and red all the other stars, only considering stars with 15<T1<17.5 in order to avoid confusion with HB/AGB stars and increasing errors at fainter mags, where the SGB is also substantially less vertical than the RGB, as well as confusion with upper MS binaries. We use this same color coding for these stars in the other CMDs. Note that the results of the addstar experiments mentioned above ensure that this magnitude range is totally unaffected by incompleteness.

The 2 CMDs using C-T1 are the clearest in demonstrating this blueward color spread along the lower RGB. The C vs C-T1 CMD also maintains the perfect separation of blue and red stars. The C-T2 CMD, although not as clearcut as in the C-T1 versions, still shows a very significant spread, and that most of the blue stars remain scattered to the blue of the main RGB locus. However, in the T1-T2 CMD, the blue stars are completely mixed with the rest of the RGB.  

\begin{figure}
\centering
\includegraphics[width=0.5\textwidth]{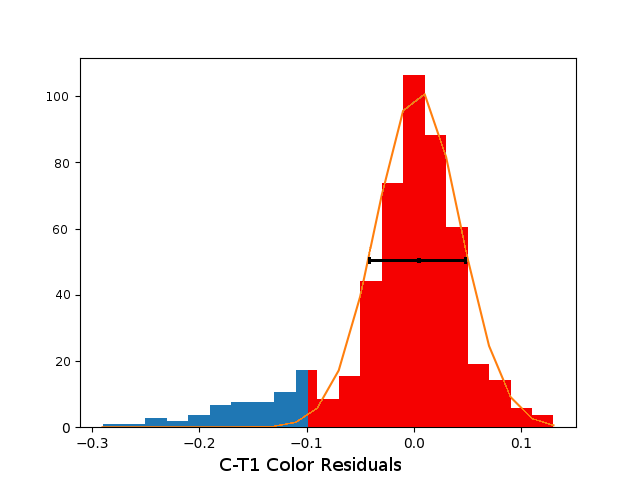}
\includegraphics[width=0.5\textwidth]{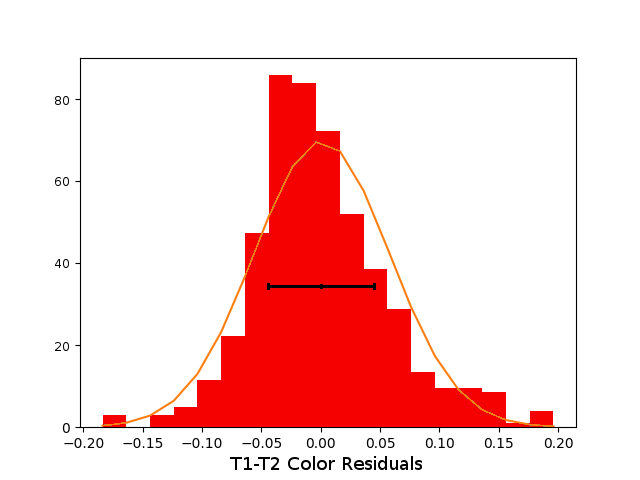}
\caption{\label{fig:hist}Color Distributions for the RGB in $C-T_1$ and $T_1-T_2$. Best fit gaussians are also shown. 
Mean photometric errors of each RGB are indicated as black horizontal lines. }
\end{figure}

 An additional test was made to investigate the significance of our finding. We fitted a fiducial curve (defined as the highest density locus of stars along the RGB) along the lower RGB in $C-T_1$ vs $T_1$ and a fiducial (using the mean of the density distribution) for $T_1-T_2$ vs $T_1$ between $T_1$ magnitudes 15 and 17.5 (the fiducials are shown in the respective CMDs in figure \ref{fig:RGB}). We then measured the color difference of a star from the fiducial and made a normalized histogram of this value. We then derived the best-fit gaussian for both histograms (Figure \ref{fig:hist}).
 
 The result was that for $T_1-T_2$ the Gaussian is centered and the sigma (0.057) is roughly comparable to that expected given the mean photometric errors (0.044) indicated as a black line at the middle of the histogram. But for $C-T_1$, there is a long blue tail, making it impossible to fit a good Gaussian. 

So we eliminate stars with $C-T_1$< -0.1, and fit the gaussian to the rest of the sample. We obtained a sigma of 0.039, fitting perfectly with the photometric errors (0.039).

Thus, all the stars coloured in blue satisfy the requirements to be considered as a different population from those well fit by the gaussian.

All this evidence suggests that we have a small bluer population in C-T1 different from the main population on the RGB. 
These results are in agreement to those found by \citet{Piotto2015}, showing a significant spread in the RGB from their magic trio data. However, it should be added that their data does not show a main locus as seen in our studies, but a homogeneous spread along the RGB.  
 
 The C filter covers the uv-blue part of the spectrum from roughly the atmospheric cutoff to 4500\AA. As discussed in \citet[see their Figure 1]{Cummings2014}, this wide range includes a number of CN bands as well as the CH band. As shown by \citet{Sbor2011}, the net effect on the fluxes of typical first and second generation red giants with otherwise identical parameters is excess absorption in the CN bands and slightly less absorption in the CH band of the SG stars compared to FG stars. This is due to their higher N and slightly lower C abundances, which creates stronger CN bands and a slighty weaker CH band. Thus, the net effect should be reduced flux in the C filter of SG stars compared to FG, making them redder. We thus expect the blue RGB stars are representative of FG and the red RGB stars SG.
We estimate from Figure \ref{fig:hist} that the blue tail is only about $14\pm 6\%$ of the main RGB population in this magnitude range.
\citet{Milone2017} derive a value of 38\% for FG stars in NGC7099 from their magic trio analysis. In fact, they find that this percentage varies tremendously over their large sample of GCs, ranging from about 10 to 70\%. They find the fraction of FG stars anticorrelates with cluster luminosity and mass but no other global properties. Given NGC7099's $M_V$ value of -7.45 \citep[updated as in 2010]{Harris2010}, the \citet{Milone2017} trend suggests it should have a FG fraction of about $35\pm10\%$, while its mass of $1.6x10^6M_\odot$ \citep{Vande2009} would suggest a FG fraction of about $15\pm10\%$ if it were to lie on the Milone et al. relations. Thus, our value is in reasonable agreement with these indications, although it is a bit lower than expected, especially compared to the value that Milone et al. derive for this GC. Note that \citet{Carretta2009} find that FG stars generally make up roughly 1/3 of the current total population, again in rough agreement with our finding.

\subsection{Radial Distributions}
 There is an intense debate about whether the different populations have significantly different radial distributions or not. Theoretically, it is generally believed that any later star formation episode should occur in the central gravitational potential where the ejecta should accumulate, so that second generation stars should be more centrally concentrated than their first generation cousins, at least initially, i.e. \citet{Dercole2008}, \citet{Carretta2010}. Subsequent stellar dynamical effects should slowly wash away this initial difference. Observationally, the evidence generally supports this: In 47 Tuc, a widely studied cluster, \citet{Milone2012} found unambiguous evidence indicating that the CN-strong population(second) is more centrally concentrated than the CN-weak(first) population. But there are exceptions: In NGC 1851, while \citet{Zoccali2009}, \citet{Marino2014} and \citet{Carretta2011} argue that different populations on the RGB have differing radial distributions, with the redder population more centrally concentrated, \citet{Milone2009} and \citet{Olszewski2009} do not find any evidence to support this.
 \citet{Cummings2014} did not find a meaningful difference in the radial distribution of their different RGB populations(p-value=0.55) in NGC 1851, but they recognized that their small sample limits the analysis, while a larger sample on the MS gave them a very significant difference in the radial distributions (p-value=0.0), indicating a redder population more centrally concentrated, and supporting \citet{Zoccali2009},\citet{Marino2014} and \citet{Carretta2011} findings. 
 More evidence indicating different radial distributions are found in \citet{Lardo2011} who studied 9 GCs, finding significant color spreads and a more concentrated redder population in 7 of them (the remaining 2 did not show radial differences between populations), while \citet{Johnson2012} found different radial distributions in NGC 6205, with the redder population again being more centrally concentrated. \citet{Dalessandro2014}, found a fully spatially mixed FG and SG in NGC 6362, which is the first evidence found of such behaviour in GCs.
 
\citet[see their Figure 12]{Carretta2009} made a plot of their [O/Na] ratios for all stars observed in their 19 clusters as a function of the distance from the cluster centre. For NGC 7099 the radial distributions show a SG generally more centrally concentrated than FG stars. However their sample was small (29 stars), limiting the conclusions.

\begin{figure}
\centering
\includegraphics[width=0.5\textwidth]{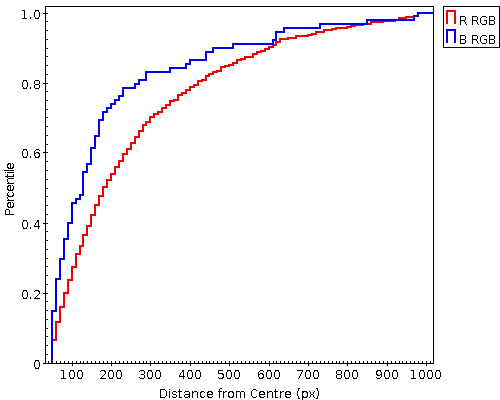}
\caption{\label{fig:radist}Radial distribution comparison between the different populations on the RGB in NGC 7099.}
\end{figure}

 Figure \ref{fig:radist} is a cumulative distribution showing a comparison of the NGC 7099 radial distribution between the bluer and redder RGB. By looking at the left histogram in figure \ref{fig:hist}, we defined $C-T_1 = -0.1$ as the limit between bluer and redder RGB stars and divided the sample in two different color subsets. Clearly, our data for these two populations in NGC 7099 shows that the blue population is more centrally concentrated than the red population. Since the blue stars are expected to be FG, this of course is opposite to the typical case where the FG stars are generally less centrally concentrated.
 A Kolmog\'orov-Smirnov test was made to find if this difference in the radial distribution between these two populations was significant or not. We found a very statistically significant difference $(p-value=0.000)$, meaning that the distributions are intrinsically distinct at essentially the 100\% confidence level.
 These results are a major challenge to most scenarios like \citet{Dercole2008} and \citet{Carretta2010}, that basically explain SG stars forming in the center of the GC while FG stars are distributed over the whole GC.

\begin{figure}
\centering
\includegraphics[width=0.5\textwidth]{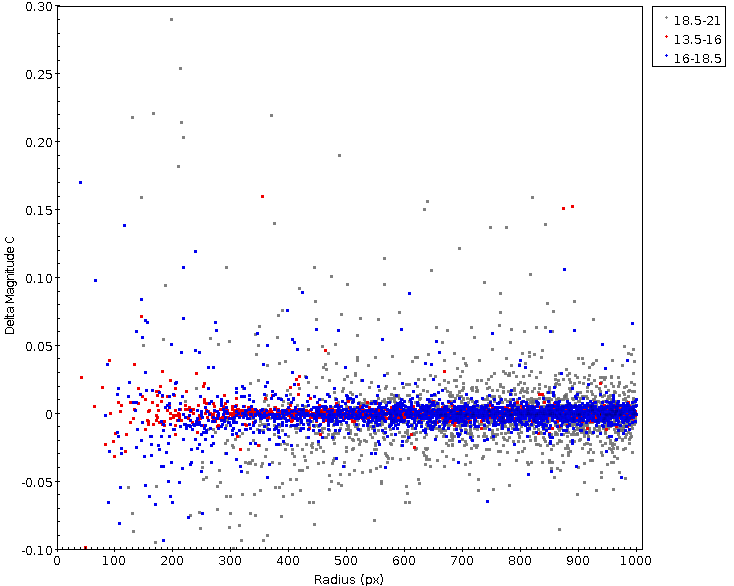}
\includegraphics[width=0.5\textwidth]{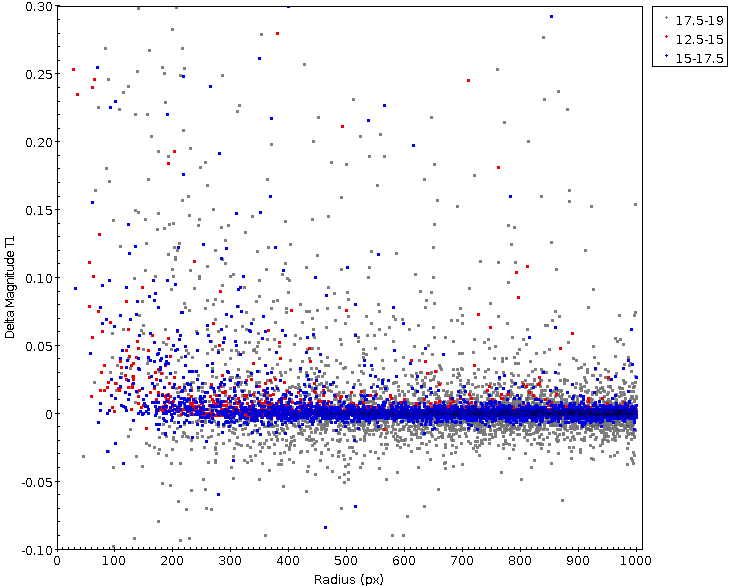}
\includegraphics[width=0.5\textwidth]{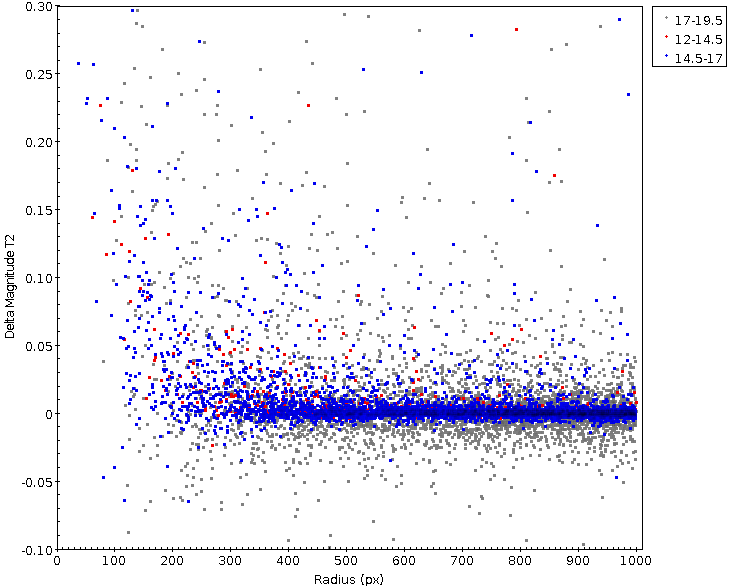}
\caption{\label{fig:blend}Input - output magnitude vs Radial distance from the center. Blue diamonds indicates stars from our subset (lower RGB) chosen to make the color distribution histograms.}
\end{figure}

One main argument that could be raised against the blue population being real is that it is centrally concentrated and lies in exactly the location where photometric blends typically lie, and the C filter has larger errors at fixed (transformed) magnitude than the other filters.  To prove that the blue population is real, we took artificial stars from our addstar experiment and made plots of the input - output magnitude for each filter vs radial distance from the center of the cluster(Figure \ref{fig:blend}). For each filter we took the range in magnitude chosen to make our color distribution histograms(Figure: \ref{fig:hist}) as a binsize ($15<T_1<17.5$, $16<C<18.5$, $14.5<T_2<17$) and create 2 more bins of the same size using magnitudes brighter and fainter than our main bin. Figure \ref{fig:blend} shows that the innermost part of the cluster does not have a bias in the recovered colors due to crowding.

 Our results seriously defy most proposed MP formation scenarios, generating the discussion presented in the next section. 
 Unfortunately \citet{Piotto2015} didn't make a radial analysis of NGC 7099. At present, no other photometric investigations of the radial behaviour of MPs in this cluster have been made. Such studies would be of great interest to see if they confirm or negate our results.

\section{Discussion and conclusions}

In the previous sections we have confirmed the utility of the Washington C filter for studying MPs. As first shown by \citet{Cummings2014}, this filter is very efficient at detecting MPs in the first test case - NGC 1851, showing MPs based on photometry obtained from a 1m ground-based telescope. We have used the same telescope (supplemented with SOAR data) to investigate a second cluster - NGC 7099. Our investigation finds that this cluster also contains MPs, confirming previous photometric and spectroscopic work. This reinforces \citet{Cummings2014} findings that MPs can indeed be studied effectively and efficiently using the Washington system. The system should prove to be especially valuable in the near future when, after the imminent demise of HST, we will be left UV-blind, so that, e.g., 2 of the "magic trio" of UV HST filters, used by \citet{Piotto2015} to study in great detail MPs in a variety of GCs, will not be available. 

We found a real spread along the lower RGB in all the CMDs that included the C filter, with $C-T_1$ showing most clearly both populations, $C-T_2$ still differentiated them substantively, but $T_1-T_2$ showed no significant intrinsic spread, mixing both populations in the RGB. The spread displayed in $C-T_1$ and $C-T_2$ was manifested by a significant number of stars lying bluer of the main red giant locus, with a color spread much larger than that expected from the photometric errors, which were well determined from addstar experiments. All this evidence indicates the presence of MPs in NGC 7099, agreeing with previous photometric and spectroscopic studies. Interestingly, we find that the bluer population, expected to be FG stars, is only about 14\% of the total population of the cluster, substantially lower than found by \citet{Milone2017} for this GC. However, our most intriguing finding is with respect to the radial distribution of the MPs. 
The radial distribution of both  populations is signficantly different, but, contrary to expected results, namely that the second population should be more concentrated, our evidence shows a more concentrated first population instead. We found that crowding cannot account for this result. This fact, as noted in the last section, contradicts most of the actual formation scenarios for MPs \citep{Dercole2008,Carretta2010}, which, as noted above, all have some fatal flaw in any case. 

 Is there a plausible theoretical explanation? \citet{Valcarce2011} presented a possible formation scenario for NGC2808, which has at least three populations. This cluster has a mass of $1.4\times 10^6 M_\odot$ \citep{Boyles2011}, a very similar mass to NGC 7099, $1.6\times10^6 M_\odot$ \citep{Vande2009}. Thus, it is reasonable to posit that NGC 7099 might have had more than 2 populations initially. Looking at the CMD of NGC2808 from \citet[Fig. 4 in their paper]{Piotto2015} and following the formation scenario for NGC2808 in \citet{Valcarce2011} we hypothesize that our bluer population being more concentrated suggests that it was possibly formed after the main redder(second) population, and thus it may correspond then to the \citet{Valcarce2011} purported third population. One is then left needing to explain what happened to the actual first, original generation. It is of course very unlikely that any purported first generation has disappeared completely. The referee suggested to us another interpretation of the observations: \citet{Milone2015a} studied MPs in NGC6266 (M62) and concluded that using F390W-F814W colors, the red MS, hosting $79 \pm 1\%$ of the total MS, is consistent with a stellar population with primordial helium, and O-rich/N-poor, while the blue MS, hosting the remaining $21 \pm 1 \%$ of the total MS, is consistent with being made of He-enhanced, O-poor/N-rich stars. Since the transmissions of the Washington C filter and of the F390W HST filter are very similar, the conclusion by \citet{Milone2015a} on M62 would be consistent with our observations of M30, and allowing the possibility that the second generation is in fact bluer than the first, and thus meaning that our results are in agreement with the usual finding that the second generation is more concentrated. We find this to be a viable alternative that requires further examination to substantiate.
 
 The median relaxation time of M30 \citep[updated as in 2010]{Harris2010} is near the median for all GCs (less than a Gyr), while its core relaxation time is quite short, only a few million yrs. One would then expect that any vestiges of inhomogeneity such as a second generation of star formation occuring in the center to disappear on a similar short time scale.
 
 Note that both the \citet{Carretta2009} spectroscopic and the \citet{Piotto2015} photometric data show no strong hints of distinct populations, but instead rather continuous distributions, with no strong concentration of stars along the sequence. Our data on the other hand do suggest a preponderance of redder stars with a scattering to the blue. Their data is consistent with NGC 7099 having canonical first and second generation stars.
 
 Our result adds to the growing amount of data characterizing MPs. Unfortunately, as noted above, no current theory is able to explain satisfactorily the wide variety of behavior that MPs display. Our result indeed increases this complexity.

Finally, we conclude that more studies in NGC 7099 and other GCs are needed. More detailed spectroscopic and photometric studies of NGC 7099 would help to clarify our results. The analysis of radial distributions in other GCs should continue to be explored to see if any other clusters display the same behaviour. Future work to search in our photometry for these multiple populations in both NGC 7099's subgiant branch and MS may help answer some of the remaining questions.

\section*{Acknowledgements}

We acknowledge Las Campanas Observatory and SOAR telescope. D.G. gratefully acknowledges support from the Chilean BASAL Centro de Excelencia en Astrof\'isica y Tecnolog\'ias Afines (CATA) grant PFB-06/2007. REC acknowledges funding from Gemini-CONICYT for Project 32140007. F.M. gratefully acknowledges the support provided by Fondecyt for project 3140177. C.M. is supported  by CONICYT (Chile)  through Programa Nacional de Becas de Doctorado  2014  (CONICYT-PCHA/Doctorado Nacional/2014-21141057). S.V. gratefully acknowledges the support provided by Fondecyt regular n. 1170518. We also acknowledge the referee for his/her many helpful comments and interesting suggested alternative interpretation of our results.

%%%%%%%%%%%%%%%%%%%%%%%%%%%%%%%%%%%%%%%%%%%%%%%%%%



%\bibliographystyle{mnras}


% Don't change these lines
\bsp	% typesetting comment
\label{lastpage}
\end{document}